\documentclass[journal]{IEEEtran}

\usepackage{longtable}

\usepackage{booktabs,tabularx}
\usepackage{supertabular,booktabs}
\usepackage[utf8x]{inputenc}
\usepackage{cite}
\usepackage{enumerate}
\usepackage{algorithmic}
\usepackage{graphicx}
\usepackage{textcomp}

\usepackage{url}

\usepackage{breakurl}
\usepackage[breaklinks]{hyperref}
\usepackage{multirow}
\usepackage[table]{xcolor}

\hypersetup{
  colorlinks   = true,
  urlcolor     = blue,
  linkcolor    = red,
  citecolor   = blue
}

\usepackage{soul}

\def\BibTeX{{\rm B\kern-.05em{\sc i\kern-.025em b}\kern-.08em 
    T\kern-.1667em\lower.7ex\hbox{E}\kern-.125emX}}
    
\newcommand{\sectopic}[1]{\vspace*{0em}\par\noindent{\textit{\bfseries #1}}} 

\newcommand{\R}[1]{\textcolor{black}{#1}}

\begin{document}


\title{Internet of Things in Space: A Review of Opportunities and Challenges from Satellite-Aided Computing to Digitally-Enhanced Space Living}

\author{Jonathan Kua,~\IEEEmembership{Member,~IEEE,}
        Chetan Arora,
        Seng W. Loke,~\IEEEmembership{Member,~IEEE,} \\
        Niroshinie Fernando 
        and Chathurika Ranaweera,~\IEEEmembership{Member,~IEEE}
\thanks{The authors are with the School of Information Technology, Deakin University, Geelong, 3220, Australia (e-mail: \{jonathan.kua, chetan.arora, seng.loke, niroshinie.fernando, chathu.ranaweera\}@deakin.edu.au).}}

\maketitle

\begin{abstract}

Recent \R{scientific and technological} advancements driven by the Internet of Things (IoT), Machine Learning (ML) and Artificial Intelligence (AI), distributed computing and data communication technologies have opened up a vast range of opportunities in many scientific fields -- spanning from fast, reliable and efficient data communication to large-scale cloud/edge computing and intelligent big data analytics. Technological innovations and developments in these areas have also enabled many opportunities in the space industry. The successful Mars landing of NASA's Perseverance rover on \R{February 18, 2021} represents another giant leap \R{for mankind} in space exploration. \R{Emerging} research and developments of \R{connectivity and computing} technologies \R{in IoT for} space/non-terrestrial environments is expected to yield significant benefits \R{in the near future}. This \R{survey} paper presents a broad overview of the area and provides a look-ahead of the opportunities made possible by IoT and space-based technologies. We first survey the current developments of IoT and space industry, and identify key challenges and opportunities in these areas. We then review the state-of-the-art and discuss future opportunities for IoT developments, deployment and integration to support future endeavours in space exploration.

\end{abstract}




\section{Introduction}

\IEEEPARstart{S}{pace} exploration has fascinated humans for centuries. With technological advancements in astronomy, satellites, telecommunications and computing capabilities, the engineering of robust and re-usable rockets and spaceships, space exploration is no longer improbable and limited to large national government initiatives. There are already many exciting initiatives developed by small startup companies to large private aerospace companies, and it is an active area of research in academia and industry alike. 

\R{On July 11 and 20, 2021, Virgin Galactic founder Richard Branson and Blue Origin founder Jeff Bezos travelled into space respectively, opening up possibilities for commercial space travel in the near future\footnote{\url{https://www.revfine.com/space-tourism/}}. Elon Musk's SpaceX has also recently become the first private company to launch a spacecraft to the International Space Station (ISS).}
\R{Earlier, the news} of two astronauts returning from the first commercially built and operated spacecraft, SpaceX’s Crew Dragon~\cite{NASA:2020aa} has generated a lot of excitement in the community.  
In the growing market of space tourism, some organisations are even building hotels in space such as the Aurora Station by Orion Span\footnote{\url{https://www.orionspan.com}} which has already sold out the first four months of travel in 2021~\cite{Orion:2018aa}. Bigelow Aerospace\footnote{\url{http://bigelowaerospace.com/}}, a space technology startup company, has been building inflatable space modules to work in low-Earth orbit, which can be used as space hotels~\cite{Galeon:2018aa}. \R{All these exciting developments in the space industry culminated with the successful} touchdown of NASA's Perseverance rover on Mars \R{on February 18, 2021} (\R{as part of the} NASA's Mars 2020 mission\footnote{\url{https://mars.nasa.gov/mars2020/}}) and the first-ever helicopter flight on Mars, followed by China's landing of the Zhurong rover on Mars on May 14, 2021.\footnote{https://www.nature.com/articles/d41586-021-01301-7} \R{These successful Martian endeavours are} indicative of recent directions in space initiatives\footnote{https://www.nasa.gov/perseverance}.

Preparations for space and interplanetary travel spans decades. For example, an 18-year old \R{has been training} to be the first human on Mars~\cite{Curtis:2019aa}, possibly in 2030, in line with NASA's mission to journey to Mars~\cite{NASA:2015aa}. Popular media also often discussed the ``Mars Race'' or ``Race to Mars'' - the competition between national space agencies and aerospace manufacturers for launching a crewed mission to Mars involving human and/or robot travel~\cite{Godd:2020aa}. \R{Furthermore}, several companies are also working towards reusable launch vehicles that can transport objects or people into orbit and return safely, thereby enabling huge reductions in cost (per kilogram) of transportation into space~\cite{Virgin:2020aa, SpaceX:2020aa}. The reduction in launch costs by a factor of as much as twenty and the ability to reuse rockets or part thereof will help drive down costs in space travel with implications for the space industry including, space tourism, power generation, development of materials, pharmaceutical research, communications, earth imaging and national security~\cite{Alvarez:2016aa, Cobb:2019aa}.

Space technologies are not new. Traditional popular space applications include the use of satellites to support telecommunications and Global Positioning Systems (GPS), research experiments in space,\footnote{Some examples at \url{https://singularityhub.com/2018/12/05/research-in-zero-gravity-6-cool-projects-from-the-international-space-station/}}, and also to support wide-area communications. With movements towards democratising access to space~\cite{Ekblaw:2019aa} and allowing commercial interests to fuel developments in space-related applications, there are increasing possibilities for space travel for a wider population and a wider group of people can influence and play a part in shaping the future of space. However, technological and regulatory challenges for space abound~\cite{10.3389/frspt.2020.00001}.

The shaping of the future of space will involve new information communication technologies (ICT) involving today's collection of technologies often termed as the Internet of Things (IoT), with \R{connectivity} technologies increasingly being combined with \R{computing and large-scale data processing capabilities powered by Artificial Intelligence (AI) and Machine Learning (ML) technologies}. For instance, one of the key visions of \R{the} fifth/sixth generation (5G/6G) \R{mobile networking technologies} is to realise a Space, Air, Ground Integrated Network (SAGIN)~\cite{Kodheli:2020aa, Liu:2018aa}. \R{The scope of IoT is very broad, and it} often refers to a collection of technologies, and a whole range of research areas\footnote{See the topics and papers published in peer-reviewed IoT journals such as ACM Transactions on IoT, IEEE IoT Journal, and Elsevier's IoT journal.}. Some active research areas in IoT include:

\begin{enumerate}[(i)]
\item Engineering the future Internet, involving 5G/6G mobile technologies, where greater support for reliable machine-to-machine (M2M) communications is required;
\item IoT infrastructure (comprising sensors, sensor networks and actuators), and associated data processing and management schemes;
\item IoT hardware technologies for low-power and long-range data transmission and processing;
\item Localisation technologies for humans, robots and ``things'' including satellite-based methods;
\item Middleware and platforms for IoT applications
\item IoT standards and semantic technologies;
\item Socio-technical issues of security and privacy, and ethics in relation to the deployment, use and business of IoT devices and services.
\end{enumerate}

As IoT technologies and deployment become more widespread and easily accessible in terrestrial environments, there are already companies using satellites for IoT-based networking applications and wide-area networking, e.g., Myriota\footnote{\url{https://myriota.com}} and SpaceX Starlink\footnote{\url{https://www.starlink.com}}. So we can start asking the question of how IoT technologies can be seamlessly integrated with existing space-related technologies to create new opportunities for future space exploration and travel -- ``Internet of Things in Space", especially in view of the developments in next-generation mobile communications, where M2M communications, Machine-Type Communication (MTC) for low-latency mission-critical applications \R{are expected to} play an important role.

In this paper, we aim to review current developments under the umbrella of ``IoT in Space", \R{spanning} IoT applications enabled by in-space technologies, current developments in satellite communications and \R{edge/fog/cloud} computing, and potential developments on IoT for new (and more futuristic) space applications. This paper is not an exhaustive review of all developments in IoT and space-related technologies, rather it serves as a vision paper that aims to explore the possibilities of \R{leveraging IoT (and associated technologies) for future space exploration endeavours.}

\R{The novelty of this review is an overview of current efforts and the identification of the unique research challenges and directions in IoT for space technologies. In particular, we provide a big picture of the myriad networks involved, from terrestrial to space for supporting IoT in space applications,  including a snapshot of current satellite-related IoT and concepts of edge computing in space. The paper is also unique in pointing out the role of IoT in emerging developments and challenges in space, including space situation awareness, inter-connecting space vehicles, and wearable IoT for space.}

The rest of the paper is structured as follows. Section~\ref{current_dev_challenges} discusses \R{the current state-of-the-art at the intersection of IoT and space technologies, i.e., satellite communications-aided IoT applications, IoT satellite-terrestrial integrated networks, satellite-based 5G (and beyond) networks for IoT, architectures/protocols, and edge computing to support IoT in space. Section~\ref{emerging_dev_challenges} presents emerging developments and challenges in the field, i.e., smart architecture and construction in space, data centres and data management in space, robots in space, connected automated space vehicles, networked wearables and applications in space, situational awareness, space debris, and traffic management in space, colonising planets, and emerging advances in satellite communications. Section~\ref{Conclusions} concludes the paper. Table~\ref{tab:acronyms} tabulates the acronyms used in this paper, and Figure~\ref{fig:taxonomy} presents the taxonomy of the topics covered in this survey paper.}

\begin{table}
\caption{\R{Acronyms and their definitions used in this paper}}
\label{tab:acronyms}
\begin{tabularx}{\linewidth}{p{1.5cm} | p{6cm}}
\toprule

\textbf{Acronyms} & \textbf{Description}\\
\midrule
AI & Artificial Intelligence\\
AR & Augmented Reality\\ 
AS & Autonomous Systems\\ 
BBR & Bottleneck Bandwidth and Round-trip time\\
BSON & Binary JSON\\ 
CBOR & Concise Binary Object Representation\\
CoAP & Constrained Application Protocol\\
CNES & National Centre for Space Studies (French: \textit{Centre national d'études spatiales})\\
DTN & Delay Tolerant Networking\\ 
ES & Earth Stations\\ 
ESA & European Space Agency\\ 
GEO & Geostationary Earth Orbit\\
GPS & Global Positioning System\\
GPU & Graphics Processing Units\\ 
GSL & Ground Satellite Links\\
HAP & High Altitude Platforms\\
ICN & Information Centric Networking\\
IETF & Internet Engineering Task Force\\
IoT & Internet of Things\\
IMU & Inertial Measurement Units\\
IP & Internet Protocol\\ 
ISS & International Space Station\\
ISL & Inter-Satellite Links\\ 
JSON & JavaScript Object Notation\\ 
LAP & Low Altitude Platforms\\
LEO & Low-Earth Orbit\\
LoRa & Long Range\\ 
LPWAN & Low Power Wide Area Network\\ 

M2M & Machine-to-Machine \\
MAC & Medium Access Control\\
MAP & Mid Altitude Platforms\\
MEC & Mobile Edge Computing\\ 
MEO & Medium-Earth Orbit\\
ML & Machine Learning\\
MIMO & Multiple Input Multiple Output\\ 
mmWave & Millimetre-wave\\
MPTCP & Multi-path TCP\\
MTC & Machine Type Communications\\
MQTT & Message Queuing Telemetry Transport\\

NASA & National Aeronautics and Space Administration\\
NB-IoT & Narrowband-IoT\\ 
NDN & Named Data Networking\\
NFV & Network Function Virtualisation\\
OASC & Optimal AS Communication\\
OEC & Orbital Edge Computing\\ 
OPF-SN & Optimal Packet Forwarding of SN\\ 
OPF-TN & Optimal Packet Forwarding of TN\\
ORM & Optimal Resource Management\\ 
OSC & Optimal Spectrum Control\\ 
OMNI & Operating Missions as Nodes on the Internet\\ 
PTO & Proximal Terrestrial Offloading\\ 
PULSAR & Prototype of an Ultra Large Structure Assembly Robot\\
QoS & Quality of Service\\
QUIC & Quick UDP Internet Connections\\
RTO & Remote Terrestrial Offloading\\
SBO & Satellite-Borne Offloading\\ 
SAGIN & Space, Air, Ground Integrated Network\\
SDN & Software Defined Networking\\
SN & Space Network\\
SINR & Signal-to-Interference-Ratio\\ 
STIDC & Space and Terrestrial Integrated Data Centre\\
STIGW & Space and Terrestrial Integrated Gateway\\ 

TCP & Transmission Control Protocol\\
TN & Terrestrial Networks\\ 
UAV & Unmanned Aerial Systems\\ 
UDP & User Datagram Protocol\\
VR & Virtual Reality\\ 

\bottomrule
\end{tabularx}

\end{table}

\section{Current Developments and Challenges} \label{current_dev_challenges}

In this section, we first provide a background and overview of satellite communications IoT (Section~\ref{sec:overview}).
We then review and discuss four key areas of current developments: (i) satellite communications-aided IoT applications, (ii) IoT satellite-terrestrial integrated networks, (iii) Satellite-based 5G (and beyond) networks for IoT, (iv) Architectures and protocols, and (v) Edge computing to support IoT in space. 

\subsection{Overview}~\label{sec:overview}
\begin{figure*}[!t]
\centering
\includegraphics[width=2\columnwidth]{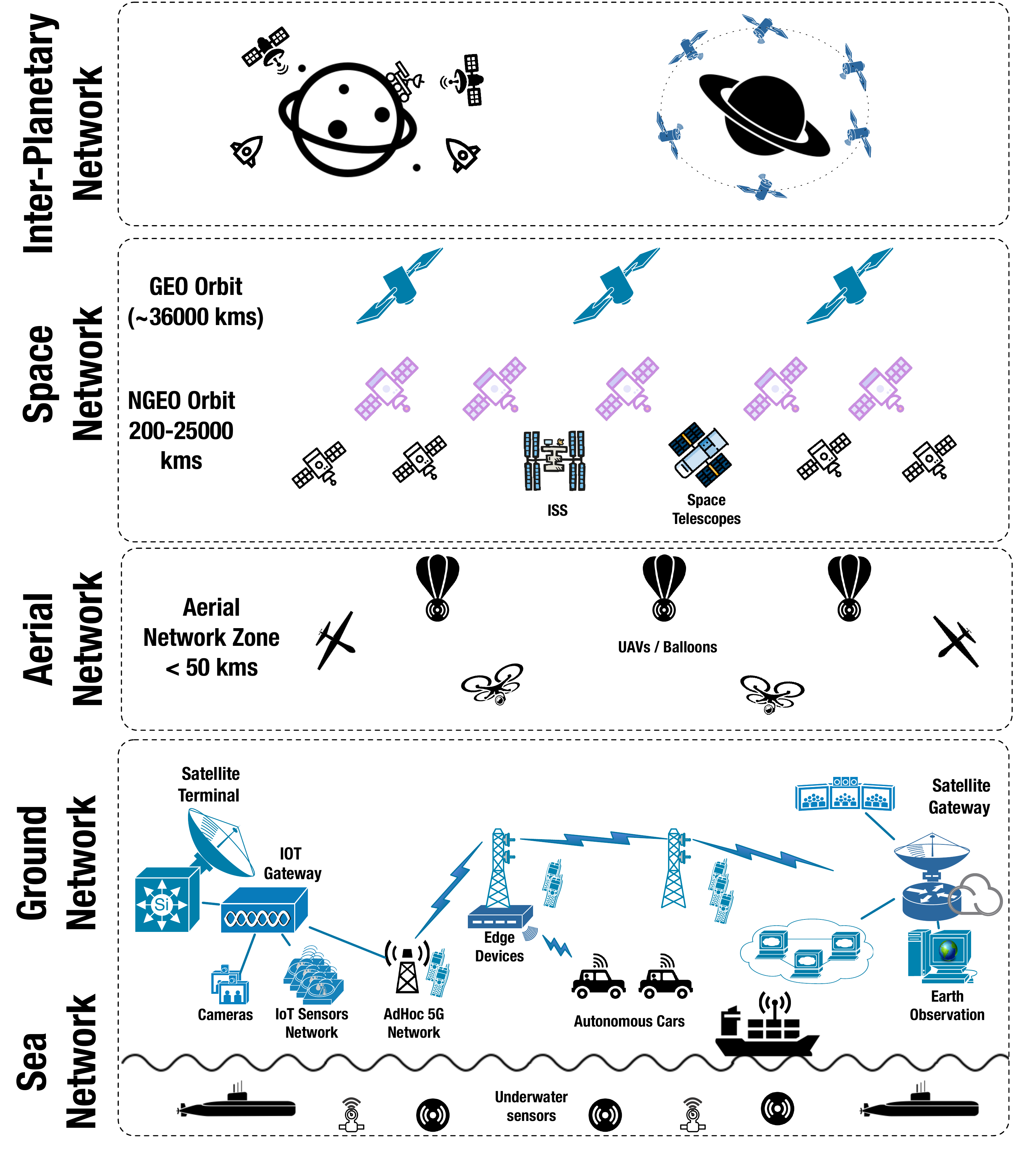}
\caption{IoT communication ecosystem spanning inter-planetary, space, aerial, ground and sea networks.}\label{fig:Overview}
\end{figure*}

\begin{figure*}
\centering
\includegraphics[width=2.2\columnwidth]{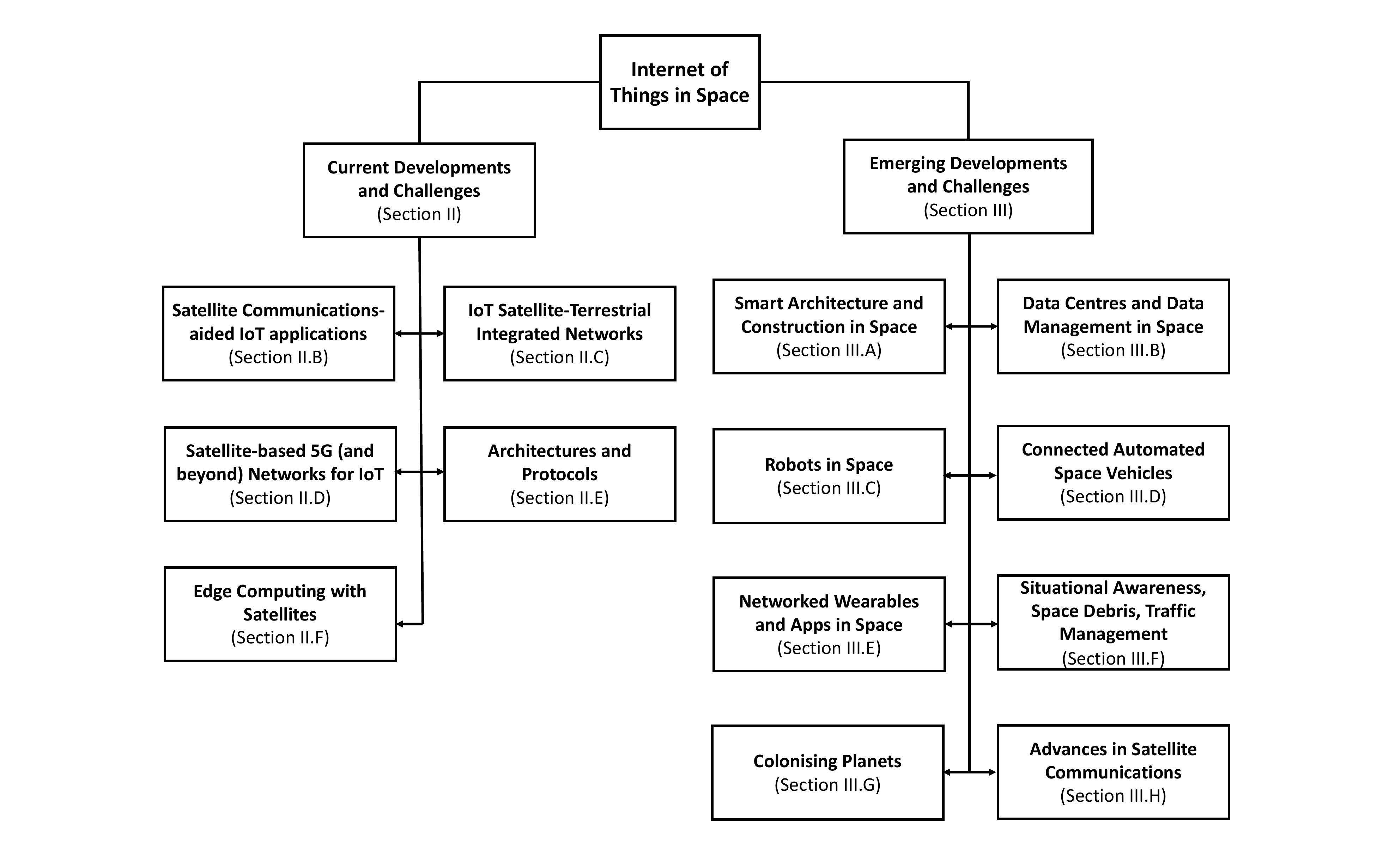}
\caption{Taxonomy of the topics covered in this survey paper.}\label{fig:taxonomy}
\vspace*{-1.5em}
\end{figure*}

Figure~\ref{fig:Overview} provides an overview of the Space and IoT communication ecosystem covered in this paper. The entire ecosystem has been broadly classified as ground network (including the underwater network), aerial network, space network and inter-planetary/outer-space network. We note that our focus in this paper is on IoT opportunities in space, and only the relevant details of the Figure~\ref{fig:Overview}'s ecosystem have been discussed in this section.   

\sectopic{Inter-planetary} communication as of today is enabled by deep space networks, such as NASA's Space Communications and Navigation (SCaN) program~\cite{Cornwell:2017aa}. The SCaN network functions over three key ground stations on earth -- California, Madrid and Canberra. Currently, the spacecrafts communicate with the deep space networks using large deep-space antennas (up to 70-metre antennas) working in higher frequency bands, such as Ka- or X-bands. The spacecrafts are typically studded with sensor and communication systems to capture inter-planetary mission data. The inter-planetary network would typically be composed of earth ground stations in SCaN (or any other deep space network in the future), orbiters on other planets, such as Mars (as depicted in Figure~\ref{fig:Overview}), rovers and any other vehicle on the surface of the planet. The rovers or spacecrafts with their sensors and communication systems can relay the information to the orbiters, which can further transmit the data to the earth ground stations. One of the main issues in inter-planetary communications is free space loss, as we discuss in the next subsection, especially in the downlink of data from outer space back to the earth ground stations~\cite{Kodheli:2020aa}. Due to the sheer distance of other planets from the earth, the power required for sending data is large, and while earth ground stations can use large amounts of power for uplink, the reverse is not true during the downlink.

\sectopic{Space network} consists primarily of the satellite communication network, with the first satellite launched in 1957.
The satellite networks can be broadly classified based on the distance the satellites orbit the earth, i.e., geosynchronous and non-geosynchronous orbits.
\emph{Geosynchronous earth orbit satellites} are placed at an altitude of $\approx$36000 kms above the earth's surface. The satellite orbit period is equivalent to a sidereal day, i.e., 23 hours 56 minutes and 4.1 seconds. The satellites in circular geosynchronous orbits and directly above the equator are known as Geostationary Earth Orbit (GEO) satellites. GEO satellites remain visible at all times from a single fixed location on earth. Therefore, the ground antennas communicating with GEO satellites do not require much adjustments once pointed, and further the receivers do not need to account for any \emph{Doppler compensation}~\cite{Elbert:08}. Hence, less sophisticated receivers can be used to communicate with GEO satellites. The main applications of GEO satellites are media broadcasting, backhauling, and communication services. In the context of IoT applications, GEO satellites are best suited for broadcasting and multi-point distribution applications. A major disadvantage of GEO satellites for IoT applications is the propagation delay of 125ms. For end-to-end two-way communication, the minimum propagation delay adds up to half a second, which is not suitable for time-critical IoT applications, such as sensor networks for autonomous driving.

\emph{Non-geosynchronous orbit satellites} are placed closer to earth than geosynchronous orbit satellites. Medium-Earth Orbit (MEO) and Low-Earth Orbit (LEO) are the two most well-known non-geosynchronous orbits. MEO Satellites are placed at an altitude of 8000-20000 kms from the earth, whereas LEO satellites are closer at an altitude of 400-2000 kms. The orbital period for MEO satellites is $\approx$6~hours and for LEO satellites is $\geq$100~minutes. Non-geosynchronous orbit satellites are better suited for IoT applications as the propagation delay is significantly less than GEO satellites. However, a constellation of satellites is required to provide continuous coverage, as LEO satellites are only visible for $\approx$20 minutes from a fixed point and MEO satellites are visible for 2-6 hours. More sophisticated ground equipment is required to compensate for doppler shift, manage communication with moving satellites, and enable satellite handovers. The main applications of LEO and MEO satellites are earth observation, geolocation services and machine-to-machine communications support.

\sectopic{Aerial network} have become increasingly popular in the last decade as an intermediate layer between the ground and space networks. Aerial networks are usually served by Unmanned Aerial Vehicles (UAV) and are classified as Low, Mid and High Altitude Platforms, commonly referred to as LAPs, MAPs and HAPs, respectively. UAVs in the form of airships, tethered balloons and drones, can provide gateway for remote sensor networks, deployed on the ground network. The data from the ground sensor networks, such as environment monitoring sensor data, can be sent to this UAV gateway which is further connected to the satellite in the space network, and can relay the data to the satellite gateways or control centres. This enables wide coverage of the IoT sensor network and provides multicast opportunities. Due to the possibility of limited line-of-sight at higher altitudes and greater distances due to the curvature of the earth and the fact that HAPs might pass beyond the coverage range of ground cellular and IoT networks, connecting UAVs or UAV networks directly to the satellites is a better alternative. The UAVs can be further utilised for direct sensing of the atmosphere at the altitude, which can be sent to control centres via space networks or a network of UAVs. 

\sectopic{Ground network} or terrestrial (IoT) networks in the ecosystem consist of the satellite ground stations, i.e., the satellite gateways and user terminals, the IoT sensor networks, terrestrial telecommunications networks such as 5G networks, optical networks and wireless local area networks, edge devices, and IoT gateways. The ground network such as a remote IoT sensor network can be connected to a user satellite terminal. Satellite backhauling can be used to access sensors and devices which are beyond the reach of terrestrial telecommunication networks, thus leading to a ubiquitous IoT network. The sensor data or other content is brought to the edge of the network by the satellites. We cover more examples of IoT satellite-terrestrial integrated networks in Section~\ref{sec:integratedNetwork}.

\sectopic{Sea network} comprises of sea ships, submarines, under-water sensors, and other maritime communication equipment. One of the most important maritime use cases is the tracking of containers and transmitting the sensing data from remote marine locations to the core network. Due to the lack of 4G/5G coverage in the seas and oceans, except ports, space segment is the only feasible option for transmitting and receiving data. For latency-agnostic use cases, e.g., the offloading of content or firmware updates, the connectivity via GEO satellites is most optimal due to the ubiquitous coverage (except polar regions) of GEO beams. For use cases where low latency is a key requirement, e.g., end-to-end inter-modal real-time asset tracking, non-GEO satellites, in particular the LEO satellites are better suited.

\subsection{Satellite Communications-aided IoT Applications}~\label{sec:applications}
In various IoT applications, end devices/sensors and controllers are dispersed across a wide geographical area. In most cases, these IoT devices are connected to the controller or to the Internet using terrestrial networks if not using the aerial network. However, in some scenarios, IoT devices are unable to connect using terrestrial and aerial networks due to limited infrastructure availability. In those scenarios, satellite communication becomes the only viable option to provide connectivity for those IoT applications. 

Most IoT applications are bounded by delay requirements. Depending on the delay requirement, IoT applications can be mainly categorised into delay-tolerant applications and delay-sensitive applications~\cite{ Ding:2020aa}. Satellite communication can be used to support both types of applications when there are no other means to provide the connectivity.   

\subsubsection{Delay Tolerant Applications}
Even though delay-tolerant IoT applications have a larger tolerable delay that could range from ten milliseconds to several seconds, these applications require continuous network connectivity to transmit important information. Providing continuous communication could be problematic in some scenarios due to geographical constraints or emergency situations. In those scenarios, satellite communications can be used to support delay-tolerant IoT applications. For example, satellite communication has been used in marine applications such as water monitoring~\cite{Bean:2017aa}, asset tracking in shipping and environment monitoring which do not entail stringent latency requirements and can thus operate with a tolerable delay~\cite{Gharanjik:2018aa}.

Another fairly common scenario where satellites can be used to complement the network availability is in emergency situations such as earthquakes, bushfires and floods where first responders with wearable and environmental sensors need to communicate with the emergency management control centres~\cite{Lee:2010aa}. Audio transmission in emergency situations is delay-sensitive but the data and video transmissions can work well with a few seconds of delay~\cite{De-Sanctis:2015aa}. Further, satellite communication can be used to provided backhaul connectivity for different types of wireless base stations to provide basic communication needs in disaster recovery areas where their backhaul links are broken due to emergency situations~\cite{Berioli:2011a}. As we \R{described in Section~\ref{sec:overview}}, both GEO and non-GEO earth orbit satellites can be used to support the delay-tolerant IoT applications. 

\subsubsection{Delay Sensitive Applications}
Delay-sensitive applications have stringent time delay requirements that are critical for \R{achieving} optimal performance. For example, IoT applications such as autonomous vehicles and industry automation require communication links that can guarantee a latency of less than 2ms~\cite{3GPP:2016aa}. However, existing upper layer protocols need to be optimally re-designed to support these delay sensitive applications. In ~\cite{De-Sanctis:2015aa}, the authors discussed different types of \R{Medium Access Control (MAC)} protocols and resource allocation mechanisms that can be used in satellite-aided IoT networks. The authors also discussed the applicability of the existing protocols to support group-based communications which is one of the major requirements of IoT applications. Authors in~\cite{Fraire:2019aa} have also discussed the suitability of existing MAC and upper layer protocols for \R{direct-to-satellite} IoT communication, particularly to be used in disaster management scenarios. However, mechanisms \R{for enhancing the reliability and low-latency performance of such communication links} warrant further research.

\subsection{IoT Satellite-Terrestrial Integrated Networks}~\label{sec:integratedNetwork}
Terrestrial networks resources are limited in capacity and coverage, hence satellites, as they provide ubiquitous coverage, are particularly beneficial in remote and disaster-struck areas. Satellite networks have the ability to strategically augment the capabilities of terrestrial IoT networks. In addition to enabling backhaul from remote areas to connected regions, satellites provide broadcast capabilities that can be leveraged for firmware updates in IoT sensor networks.

IoT sensor networks produce data traffic that is sporadic and \R{the generated packets are usually small in size. Therefore,} when combined with satellites, the data traffic pattern of IoT terrestrial networks can lead to expensive signalling over the satellite link. IoT protocols such as Constrained Application Protocol (CoAP) and the Message Queuing Telemetry Transport (MQTT) were originally designed for constrained devices, without
taking the satellite communication constraints into consideration. Optimisation of IoT protocols is required to reduce the amount of traffic load over the satellite return channel, to improve the Quality of Service (QoS) of data delivery, and to avoid expensive satellite-bandwidth wastage due to the transmission of obsolete data~\cite{Giotti:18,Soua:2018aa}. Soua et al.~\cite{Soua:2018aa} performed experiments on IoT protocol optimisation as part of the M2MSAT European Space Agency (ESA) project, wherein they optimised the MQTT and CoAP protocols using caching, filtering and aggregation of the data on the edge. The experiments were performed for satellite backhauling from the remote areas to the core network using GEO satellites simulator. 

As discussed in Section~\ref{sec:overview}, GEO satellites are only suited for IoT applications where the content needs to be delivered to the edge or to the users, without much latency constraints. However, non-GEO satellites are better suited for most IoT use cases. Non-GEO satellites, in particular LEO satellite-based IoT systems, have a very dynamic traffic distribution due to the limited visibility of non-GEO satellites. It is thus particularly important to study the resource management strategy in non-GEO integrated networks~\cite{Jin:19}. Siris et al.~\cite{Siris:16} proposed an integrated satellite-terrestrial IoT network using an \R{Information Centric Networking (ICN)} architecture to overcome the aforementioned issues, and performed experiments with connectivity via LEO satellites, \R{focusing on collecting environmental data with IoT sensor networks.} The \R{authors evaluated} three models in \R{their} simulation: (i)~\emph{Message aggregation at the edge} - saving data traffic and control traffic; (ii)~\emph{Confidential data transfer}, i.e., edge aggregator cannot look into data but informs subscribers that data is available - savings control traffic; and (iii)~\emph{Individual proxy for each IoT node} - highly secure data transfer between each IoT node and a subscriber. All three models are important for the future architecture considerations of IoT integrated networks, including the integration of all layers below the space network \R{as presented} in Figure~\ref{fig:Overview}. 

\subsection{Satellite-based 5G (and beyond) Networks for IoT services} \label{sec:satellite5g}

The fifth generation (5G) mobile technology promises higher bandwidth capacities, lower latency and higher reliability for emerging time-sensitive and mission-critical IoT applications. Whilst mobile service providers worldwide are currently \R{deploying} 5G in their networks, academia and industry have already started developing a road map for the sixth generation (6G) technology which \R{is expected to be standardised and deployed in the next decade}. 6G is expected to provide \R{data rates in the order of hundreds of Gbps (most likely exceeding 1Tbps) and sub-millisecond} ultra-low latency over ubiquitous three-dimensional coverage areas~\cite{Saad:2019aa, nayak20206g, yang20196g, dang2020should, giordani2020toward, strinati20196g}.  Ground and aerial networks will be used in conjunction to achieve the key \R{performance} indicators of 6G, whereas satellite communication using LEO satellites and CubeSats will be used to complement the 3D network coverage (connecting cell sites in rural areas with the rest of the network)~\cite{Cao:2018aa}. Therefore, \R{one of the main research focus} of 5G and beyond networks \R{is the full integration of} ground, aerial and satellite communication networks.
 
Researchers have investigated techniques and architectures \R{for integrating} satellite communication into 5G (and beyond) networks to support emerging IoT services. For example, Gineste \textit{et al.} conducted a preliminary analysis on the seamless integration of satellites and HAPs into 5G networks~\cite{Gineste:2017aa}. The authors described the necessary modifications for HAPs to operate on 5G systems using Narrowband-IoT (NB-IoT). They showed that the system can operate at low bitrate through satellite components with minimum configuration update providing continuity of service while complementing terrestrial infrastructure for NB-IoT services.

In~\cite{Bontu:2014aa}, Bontu \textit{et al.} proposed an IoT wide-area communication system concept deployed within an operator's licensed macro-cellular band which is suitable for low-energy, low-complexity IoT modules with low-priority and infrequent IoT traffic. The authors also proposed a simplified air interface protocol for IoT and a simultaneous uplink IoT communication.  In~\cite{Fang:2020aa}, Fang \textit{et al.} discussed three basic cooperative models and techniques that can be used to implement hybrid satellite-terrestrial networks to support 5G mobile networks. The authors have also identified the challenges in implementing hybrid network to support 6G and provided an analysis of ongoing research in this area. In particular, the authors have discussed the importance of incorporating techniques such as \R{SDN, NFV, AI and Blockchain technologies} into satellite-based 6G hybrid networks to support emerging IoT services. 

Furthermore, researchers have also investigated the benefits and requirements that we need to satisfy when we deploy 5G and beyond networks to support IoT services using satellite links. Evans \textit{et al.}~\cite{Evans:14} delineated key requirements for deploying 5G networks delivering IoT services using satellite terminals, \R{also enabling communication resiliency and ubiquitous coverage}. Authors have also investigated new physical and data link layer protocols and architectures \R{for reducing the} energy consumption of satellite terminals.

\R{These are just a subset of the many advances in satellite communications in the era of 5G and beyond. For a more comprehensive review, we refer the readers to~\cite{Liolis:2019aa}.}

\subsection{Architectures and Protocols} \label{sec:architectures}

In this section, \R{we review existing architectures that aim to integrate the terrestrial IoT networks with space networks.}

\subsubsection{H-STIN architecture for IoT}

The Heterogeneous Space and Terrestrial Integrated Networks (H-STIN) framework was proposed in~\cite{Chien:2019aa}. H-STIN aims to integrate various system architectures and wireless communication technologies that are currently being deployed. This is to meet the requirements of massive Machine Type Communication (MTC) with high bandwidth requirements, and edge-less communications -- achieving an intelligent, rapid, and efficient framework for IoT communications with space-based technologies.

The development of H-STIN was motivated by the challenges posed by the integration of heterogeneous communication protocols, routing problems, and resource allocation at a large scale. Satellite communications have a wide range of coverage, with end devices receiving satellite signal from virtually anywhere on earth if there is no interference. Scientific explorations on vast oceans, deep space, volcanic and other environments rely on satellite communications. However, satellite communications have higher latency than Terrestrial Networks (TN) and is vulnerable to the climate such as signal attenuation resulting from clouds and rains. Direct communications between satellites and end-devices can be challenging, hence some methods use the signal forwarded through TN.

Different network architectures have different communication protocols, transmission technologies and system architectures. In many cases, these systems operate independently. For example, TN still uses many well-known, \R{traditional/conventional} protocols and network architecture such as the Transmission Control Protocol/Internet Protocol (TCP/IP) suite, Named Data Networking (NDN), Software Defined Networking (SDN), Delay Tolerant Networking (DTN) (which originates from NASA’s research from interplanetary Internet). TN may also includes Low Power Wide Area Network (LPWAN), which uses a licensed band of NB-IoT and unlicensed bands such as Long Range (LoRa) and Sigfox\footnote{\url{https://www.sigfox.com}}. In the Operating Missions as Nodes on the Internet (OMNI) project~\cite{NASA:2008ab}, NASA aimed to integrate the TCP/IP suite into satellite communication and adapt SDN to address the problem of inflexible routing. On the other hand, various technologies are used to connect various existing satellite systems in Space Networks (SN). The Internet Protocol (IP) still remains the core of two major networks.

More specifically, the H-STIN architecture aims to integrate IoT, mobile networks, and satellite networks (consisting of satellite network, backbone network, and wireless network using unlicensed/licensed bands), and leverages TN, Autonomous Systems (AS), and Self-organisation Satellite Terrestrial Integrated System (SSTIS).

The proposed framework uses SDN and \R{Network Function Virtualisation (NFV)} for complex task management in SSTIS, \R{comprising of} three distinct layers:
\begin{enumerate}
    \item {\it{Perception layer}}: Perceives network information in SN and TN, e.g., network traffic load and speed, and Signal-to-Interference-Ratio (SINR). The layer combines SDN and NFV to dynamically adjust network \hbox{resource allocation.}
    \item {\it{Cognition layer}}: Monitors network information based on data observed from the perception layer. This addresses the weather influence, e.g., rain attenuation, and enables \R{the prediction of} network traffic and environment state using data mining methods, clustering and classification methods.
    \item {\it{Intelligence layer}}: Plans route and manages resources based on the results from the Cognition layer. This addresses the NP-hard problem using meta-heuristic algorithms, machine learning and dynamic programming to find an optimal solution in a limited time.
\end{enumerate}

This architecture also introduces an intelligent data centre and associated technologies for implementation. It integrates SN and TN packet forwarding, and use the concept of service orientation to achieve highly efficient resource allocation. In TN, the backbone network is responsible for the connection of the entire network for large data transmission with high bandwidth requirements.

H-STIN also has a Space and Terrestrial Integrated Gateway (STIGW) that serves between TN and SN. The Autonomous System (AS) implements distributed control and resource management whereas the STI-Data Centre (STIDC) collects information on routing table, network speed, etc. Within an AS, the data centre provides all routing control and management. Routing is always dynamic, for example, data packets are being forwarded via LEO or MEO satellites when the distance is shorter, and forwarded via GEO satellites when the distance is longer.

The SSTIS Intelligence layer has multiple controls: Optimal Access Control, Optimal Packet Forwarding of SN (OPF-SN), Optimal Spectrum Control (OSC), Optimal Packet Forwarding of TN (OPF-TN), Optimal AS Communication (OASC), Optimal Resource Management (ORM). Each of these controls enables the coordination of data transmission across the H-STIN architecture.

One of the key challenges that arise from this work is the routing optimisation problem to support a large number of devices. The authors proposed (but not yet evaluated) Large-scale Integrated Route Planning and Large-scale resource allocation approaches. Some key technologies not discussed in the work includes using Multiple Input Multiple Output (MIMO) for SN, protocol integration, other wireless technologies. These can form a basis for further enhancements to the H-STIN architecture.

\subsubsection{CubeSats and SDN-based technologies}

Wireless communication technologies such as LPWAN, LoRa, SigFox and NB-IoT represents key enablers for IoT. However, cost-effective \R{wide area coverage} is still a challenge with significant costs associated with building infrastructure in remote areas. While LEO satellites are a potential solution for \R{wide area coverage}, problems such as long development times and high costs make them challenging to keep up with the increasing number of IoT devices.

Recently, a new class of miniaturised satellites, known as CubeSats~\cite{Saeed:2020aa}, has emerged as a viable solution towards establishing global connectivity at low costs. In~\cite{Kak:2018aa,Akyildiz:2019aa} the authors proposed a CubeSats design that supports multi-band wireless communication at microwave, millimetre-wave (mmWave), and Terahertz (THz) band frequencies. In recent years, several CubeSats-based IoT and broadband solutions such as Iridium NEXT\footnote{\url{https://www.iridium.com/blog/category/network/iridium-next/}}, SensorPOD, Astrocast\footnote{https://www.astrocast.com/}, Fleet\footnote{\url{https://fleetspace.com/}}, etc. However, low data rates and the lack of continuous global coverage remain a challenge. Hence, the authors proposed an SDN/NFV-based ``Internet of Space Things'' (IoST) to enable (i) backhaul in the sky, (ii) eyes in the sky, and (iii) cyber-physical integration.

Applications that use backhaul in the sky include remote communications, load-balancing, emergency communication services. ``Eyes in the sky'' applications involve sensing and monitoring applications such as aerial reconnaissance, asset and environment monitoring, disaster prevention. Cyber-physical integration involves the integration of localised information collected by on-premise sensors with global information sensed by CubeSats.

The authors performed a comprehensive evaluation of the proposed SDN-based CubeSats network, analysing the performance of Ground Satellite Links
(GSL), Inter-Satellite Links (ISL), next-hop metrics and end-to-end system operation, under various configurations. They investigated the impact of different orbital configurations and carrier frequencies (ranging from S-band to THz) on data rates, link latencies, next-hop availability and access duration. They also looked into a use case where data is delivered across two major cities. The authors concluded that the proposed model has potential for remote sensing, cellular backhauling and mission critical communications.

\subsubsection{Space-Air-Ground (SAG) IoT Network}

As discussed earlier in this paper, the integration of space, air and ground networks (with various characteristics in the three network segments) in the context of IoT (with AI and ML-based data processing and analytics approaches) is an active area of research~\cite{Liu:2018aa, Kato:2019aa}. \textit{Space network} typically refers to multi-layer satellite networks where satellites in different orbits will complement each other to meet QoS requirements. It uses inter-layer and intra-satellite links to establish a reliable communication network, which can be efficient for real-time communication. \textit{Air network} typically refers to LAP or HAP. UAV-IoT networks are a good example of LAP, whereas large UAVs, airships or hot air balloons are known as HAP. \textit{Ground network} refers to base stations, relay stations, vessels, mobile nodes connected via point-to-multipoint networks or mesh networks.

In~\cite{Hong:2020aa} Hong \textit{et al.} studied the potential integration of SAG by applying network slicing. In particular, they considered the effects of UAV on mmWave channels. They demonstrated preliminary simulation results on the performance of such use-case. A Cloud-based modular simulation system for 5G is proposed for future IoT research. This simulation toolkit is meant to be highly efficient and precise, and has flexible configuration options. The use of UAVs for IoT has become increasingly popular. UAVs that are equipped with communications and IoT devices can provide services such as data collection, target identification, temporary communication. They can be easy to deploy and manage, with a wide range of applications and affordable cost.

Satellite network covers a wide area and provides seamless connectivity to ocean and mountain areas, whereas air networks enhance capacity in areas with high service demand, and densely deployed ground segment systems support high data rates. The authors comprehensively studied the potential integration of space, air and ground networks with IoT-based networks. Prior work (such as the Iridium system) did not consider the integration of IoT in these systems, they are more focused on a single application use-case, such as voice or video transmission. Therefore, the authors proposed a novel SAG-IoT network paradigm by integrating satellite communications, air communications, 5G and IoT technologies. Each node in space, air, ground planes closely cooperate and coordinate to transmit messages efficiently.

Smart cities are one of the important applications of SAG-IoT. These networks can support vertical markets with different requirements (automotive, environmental protection, medical search and rescue). However, SAG-IoTs need to be flexible, manageable, scalable, customisable to support multi-service demands. Since SAG is a stereo network with multi-path propagation in mmWave signal, a good 3D channel modelling is required. Flying UAVs will have an impact on the propagation of wireless signals. Phenomenons such as signal reflection, scattering and diffraction can produce multi-path effects which will distort the wireless signals. The motion of flying UAVs can also cause frequency shifts.

Therefore, another contribution of the authors is a 3D imaging methodology to study the effects of UAVs on mmWave signal. They set up a mmWave channel modelling methodology for UAV networks, which include mmWave measurement, Doppler Shift analysis (Doppler/micro-Doppler effects caused by motions), and integrate them into the 5G measurement and simulation platform. In an IoT network where UAVs are considered, UAVs can serve as a temporary base station and a relay node to collect IoT information, such as camera sensors, etc. The two key technologies that enable this is the formation of flight and motion control of UAV groups and the communications/IoT data processing UAVs.

Another important factor to consider is the formation of multiple flying UAVs. There are two main categories of such formation:
\begin{enumerate}
    \item A centralised control approach using the ``Leader-Follower'' model where the leader controls the flight of the entire formation.
    \item A decentralised coordinated control approach where each UAV operates with its own flight path, and sends information to other UAVs to prevent \hbox{flight-path duplication.}
\end{enumerate}

The work presented in this section summarised the importance of having an integrated architecture for space, air and ground networks. A good coordination among the different ``things'' in these three network segments can ensure the efficient use of resources and better leverage the capabilities of communication technologies.

\subsubsection{Data transmission and exchange formats}

\R{The transmission of} bulk data streams between space and ground has been around for a while~\cite{Wang:2009aa,Caini2007aa,Akyildiz2001aa}. There are numerous works on optimising \R{TCP, the de facto protocol for data transmission} for network paths with large bandwidth capacities and high delays, such as in satellite networks and space-ground networks. \R{The Internet Engineering Task Force (IETF) has been active in this area, for example, RFC 2760~\cite{rfc2760} specified ongoing TCP research for satellite communications, and more recently RFC 8975~\cite{rfc8975} outlines work in network coding for satellites. More recent research also focuses on incorporating and modifying state-of-the-art transport protocols and mechanism for satellite-based communications, such as Bottleneck Bandwidth and Round-trip time (BBR), Quick UDP Internet Connections (QUIC) and Multiplath TCP (MPTCP)~\cite{kuhn2020quic,kuhn2020quic,wang2018performance,jones-tsvwg-transport-for-satellite-00}}.  

The emergence of integrating IoT in space networks have opened up possibilities to optimise various communication protocols (not just TCP) and data exchange formats (and associated textual/image compression technologies) to enable \hbox{efficient data transfer and communication.}

Due to the heterogeneity of IoT services offered by various devices across different communication technologies, seamless integration of data transfer formats remain a key challenge in IoT communication. Furthermore, adding an additional layer of satellite communication on top of well-known IoT technologies such as NB-IoT, \R{LoRaWAN}, LPWAN, IEEE 802.15.1, IEEE 802.15.4 creates additional challenges. In addition, given the expensive cost of satellite links, an approach to optimise data exchange formats can play an important role to ensure efficient data transmission.

Lysogor \textit{et al.}~\cite{Lysogor:2018aa} presented a concise survey on data exchange formats for IoT services over satellites. Following the survey, they also compared the formats by measuring their energy efficiency, applicability in remote areas, network coverage, and operation in an unlicensed frequency band. They concluded that the Iridium Short Burst Data network is best suited for IoT applications. However, it limits the message size for the transmitted data, hence emphasising the importance of selecting the \hbox{appropriate data exchange format.}

JavaScript Object Notation (JSON) is the most commonly used data format in IoT systems. For example, MQTT messages use JSON-encoded data. The textual nature of JSON allows easier debugging, troubleshooting, and having flexible key-value representation allows for flexibility, meaning most IoT applications can read and support JSON formats. Binary JSON (BSON) is a binary version of JSON, which handles binary data inside JSON messages. BSON does not contain data encoding optimisation, and is mostly similar in size when compared with JSON. \R{Concise Binary Object Representation (CBOR), as defined in RFC 7049,} use binary data representation and optimise \R{message} sizes using appropriate formats, allowing for full optimisation of the overall payload. Both of them do not require pre-defined keys before data transmission. JSONC\footnote{\url{https://github.com/tcorral/JSONC}} uses the zlib\footnote{\url{https://zlib.net/}} library for data compression of JSON data. It does not optimise the data representation. This format significantly optimises message sizes when transmitting large textual data.

Protocol Buffers use a binary format, allowing message size to be significantly reduced by using an optimised value type encoding and a pre-defined key structure. It minimises the payload by sending key identifiers instead of key names. All of the information is stored in a binary format.

\R{In summary,} data formats with pre-defined keys are more efficient for IoT data since keys can be sent once before regular value transmissions begin. The authors presented a simulation model to evaluate the efficiency of different data exchange formats. The model consists of seven different nodes, and the communication between each node uses a different data exchange format. These nodes include LoRa endpoint, LoRa gateway, LoRa-Iridium gateway, Iridium satellite system, Iridium ground gateway, Iridium MQTT gateway, and a central data collection point. Each node uses one of the data exchange formats mentioned above. Evaluation results show that Protocol Buffers is the most efficient \R{format}, transferring four times more data than using JSON.

\subsection{Edge Computing with Satellites} \label{sec:edge_computing}

Satellite-enabled Internet at the LEO level, also known as Satellite Terrestrial Network (STN), is intended to augment the existing terrestrial Internet, especially for users in areas lacking terrestrial communication infrastructure, such as in remote areas and for aeronautical and maritime users~\cite{Hu:2001aa}, in order to provide ubiquitous access to high-speed Internet around the world.

A key distinction between fog/edge computing via STNs vs conventional fog/edge computing, is the distribution of mobile users. In a conventional fog/edge computing environment, mobile users are typically densely distributed. However, in many STN fog/edge computing scenarios, the users are sparsely distributed, and typically access the Internet via a small terminal station, with minimal computation and storage capacities. However it is not economically viable to deploy a Satellite-based edge/fog server for such a small number of users. This issue is discussed in~\cite{Zhang:2019aa}, where the authors investigate methods to improve the QoS of STN mobile users via edge computing. Computation offloading in such a scenario can occur in three different ways; Proximal Terrestrial Off loading (PTO), Satellite-Borne Offloading (SBO) or Remote Terrestrial Offloading (RTO). In PTO, users offload tasks to an edge server located in the terrestrial stations, thus avoiding backhaul transmission to the satellites. In SBO, the LEO satellites themselves are equipped with edge computing capability, thus avoiding the need for communications with remote clouds and reducing traffic between satellite and terrestrial backbone networks. In RTO, the edge servers are deployed in the terrestrial backbone network gateways. While all three of the aforementioned approaches are aimed at reducing latency, the highest reduction would come from PTO, followed by SBO, and last, RTO. However, there are other constraints such as hardware costs and energy consumption that needs to be taken into consideration that can make the above three strategies less practical. In~\cite{Zhang:2019aa}, the authors propose an alternative strategy of combining the resources of multiple Mobile Edge Computing (MEC) servers within the coverage of a LEO satellite, using dynamic NFV, enabling the STNs to centrally manage task offloading. The MEC servers use a cooperative offloading scheme to complete the user tasks. Simulation results show a reduction in user-perceived delay and energy usage.

LEO satellites are often used to collect data in space for applications such as weather forecasting environment monitoring and target surveillance. Once collected, these data need to be downloaded to ground servers via Earth Stations (ES). However, due to their high speed, satellites have limited contact time with ES, which is often insufficient to download all the collected data. One approach is to optimise the scheduling of data exchanges between satellites and the ES to optimise the total throughput of data downloading, within the available time window~\cite{Spangelo:2015aa, Marinelli:2011aa}. In~\cite{Jia:2017aa}, the authors propose another approach, suited to cases where a very large amount of data needs to be transferred from the satellite to the ES. Here, satellites use ISLs to share the data \R{amongst} themselves collaboratively, prior to coming in to contact with the ES. In this way, satellites with a high data to contact time ratio can offload some of the data to other satellites with low data to contact time ratios. One of the challenges in this method is to manage the overlapping of contact windows of multiple satellites with the same ES. The proposed method uses a time-sharing method that initially allocates equal download time to all of the overlapping satellites, and then, at the end of each offload, the download time is iteratively re-adjusted, until either all satellites complete their downloads during the time window, or an optimum is reached. Simulations using two well-known LEO satellite constellations, Globalstar~\cite{Smith:1996aa} and Iridium~\cite{Pratt:1999aa} show a significant increase in the throughput of data downloading.

In~\cite{Xie:2020aa} the authors propose an architecture named Satellite Terrestrial Integrated Edge Computing Network (STECN), where a LEO satellite network works with hierarchical and heterogeneous edge computing layers and clusters to service user requests. In the STECN approach, the LEO satellites are also equipped with computational capacity, and the architecture supports content caching, computation offloading and network services. The STECN architecture comprises of the edge computing service providers at three layers; the satellite network, the terrestrial network and clusters (e.g., at locations with minimum infrastructure such as aviation, marine clusters). Users can offload to a relevant layer depending on the context. The authors suggest a cooperative and multi-node computation \R{offloading} approach between user devices, and the edge service layers as a technique to improve efficiency. However, STECN is still at the architectural conception stage and no experiments or implementations have been discussed.

Offloading algorithms in fog/edge computing via STNs need to address the intermittent terrestrial-satellite communication caused by satellite orbiting, unlike in conventional fog/edge computing. To address this challenge, the work in~\cite{Wang:2019aa} employs a method based on game theory to optimise offloading from user devices to satellite-based edge computing. In this scenario, the satellites are equipped with the computational capacity and contain the edge server on-board. Due to the intermittent connectivity, tasks can only be offloaded when the satellite is flying over. Hence, the method consists of three components; namely, the satellite's orbit model, the communication model and the computation model of task execution.

It is assumed that the size of the results transmitted from the satellite to a terrestrial-based mobile device is much smaller than the uploaded data, hence it is disregarded in the communication model. In the task execution model, it is assumed that the user devices are not cooperating and will choose the offloading strategy that offers the best QoS to each of them. Hence the computation offloading problem is formulated with game theory. Here, the offloading game considers the average response time and average power consumption of a task as performance metrics. The Nash equilibrium of the offloading strategy of each device is obtained via an iterative algorithm. Simulations using the Iridium constellation~\cite{Pratt:1999aa} show that this method can reduce the average cost of offloading substantially.

\subsection{Summary and Discussion}
\R{Table~\ref{tab:current} summarises our discussion in this section. The challenges have been highlighted in the table and current representative work have been noted.} This section first provided an overview of satellite communications and a description of the various network segments, including the inter-planetary, space, aerial (or air), ground and sea networks. \R{This is followed by a survey of current} technologies supporting delay-tolerant and delay-sensitive applications, IoT satellite-terrestrial integrated networks, and how next-generation 5G/6G mobile networks can support IoT and space-related services. We also discussed various well-known architectures such as H-STIN and other communication protocols/frameworks to deliver optimal quality of service under various constraints. Edge computing with satellites and Satellite Internet is another key aspect of delivering optimal IoT services with satellites. The question of how and where to compute and process data becomes increasingly important as latency becomes critical in many modern applications. With the advancements of technologies reviewed in this section, the community is working towards realising the goal of an integrated network that provides seamless connectivity and efficient delivery of new IoT applications and services.

\section{Emerging Developments and Challenges} \label{emerging_dev_challenges}
In this section, we discuss potential future  developments, some of which are of a more nascent nature. We consider applications and scenarios that are of a more speculative nature but attempt to extrapolate from technological developments as of today. \R{More specifically, we consider: (i) Smart architecture and construction in space, (ii) Data centres and data management in space, (iii) Robots in space, (iv) Connected automated space vehicles, (v) Networked wearables and applications in space, (vi) Situational awareness, space debris, and traffic management in space, (vii) Colonising planets, and (viii) Advances in satellite communications.}
    
\subsection{Smart Architecture and Construction in Space} \label{sec:smart_architecture}

Architecting and constructing smart buildings in space for (future) space living is an interesting avenue of research with unique design and development challenges. NASA has demonstrated a system for autonomous assembly and disassembly of an 8-meter planar structure comprising 102 truss elements covered by 12 panels as far back as 2002~\cite{Doggett:2002aa}. An effort to use robots to assemble Truss modules to form mirrors for telescopes is described in ~\cite{Lee:2016aa}. Different types of large structures can be autonomously assembled in space, in particular, structures that are too large to be launched as one piece into space. As reviewed in~\cite{Roa:2019aa}, even a large telescope can be assembled, once its components have been launched into space. The PULSAR (Prototype of an Ultra Large Structure Assembly Robot) project\footnote{\url{https://cordis.europa.eu/project/id/821858}} aims at developing technology for the on-orbit assembly of a large primary mirror using an autonomous robotic system. Remote control  of a large robotic arm for assembly is not feasible given the significant time required for the synchronisation of operator commands and actions. Hence, the autonomous performance of an (even complete) sequence of assembly tasks is required. The ISS has a specialised robotic arm for space assembly.

Robots that assemble structures have been investigated in different projects, though not specifically for space structures. A swarm of robots can construct a structure collaboratively~\cite{Nagpal:2002aa, Chandler:2019aa}. It is not only robots for on-orbit in-space assembly, but robots can be developed to assemble structures on other planets, e.g., building a moon base.

\begin{figure}[hbt]
\begin{center}
\includegraphics[width=60mm, height=50mm]{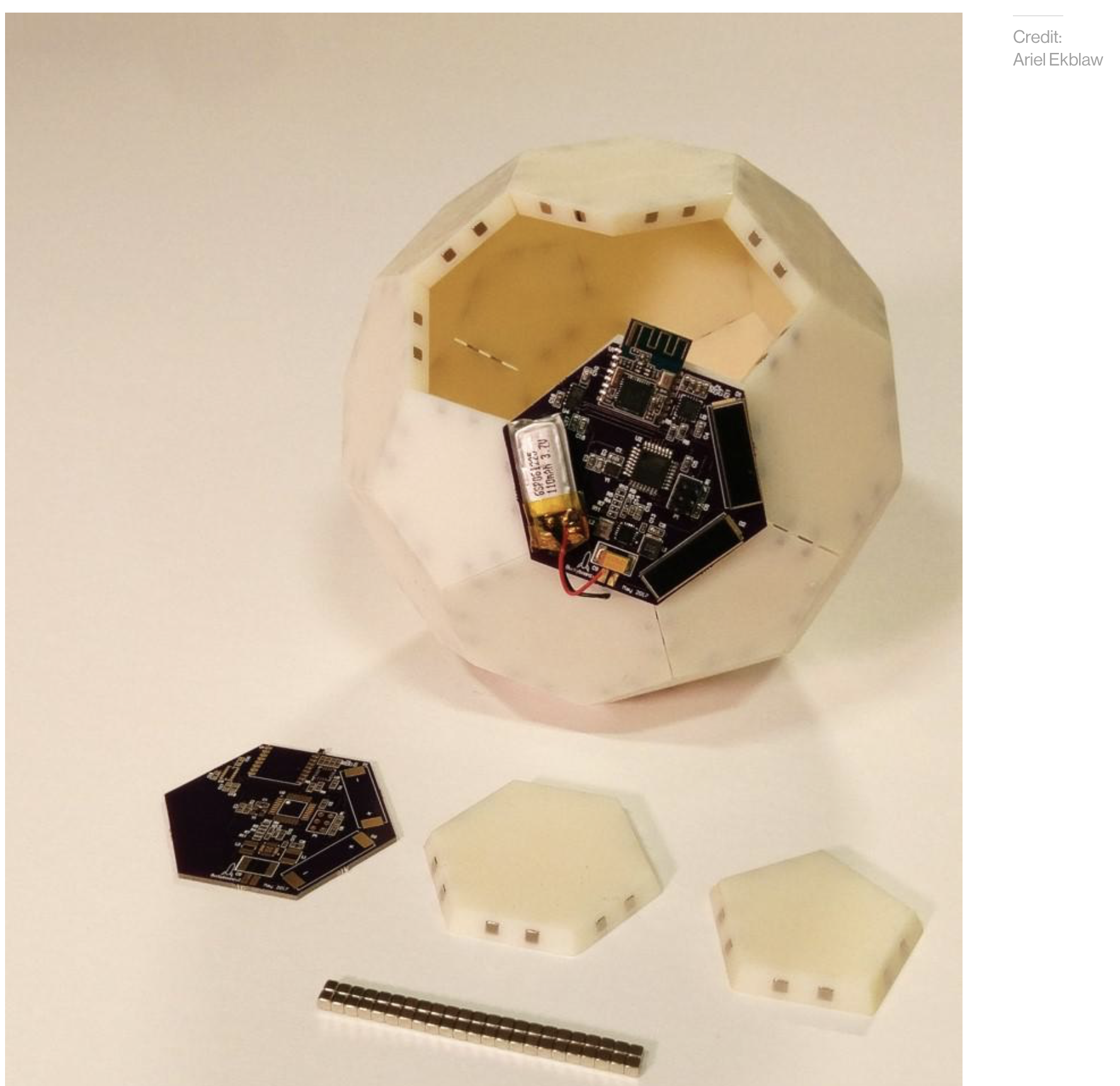}
\caption{``Bucky Ball'': A self-assembling structure in space (zero gravity)~\cite{MIT:2020aa}}
\label{bucky}
\end{center}
\end{figure}

Apart from using robots to assemble structures, there could be other interesting functions. For example, there could be self-assembling structures that form a ``Bucky Ball'' as a goal, proposed in ~\cite{Ekblaw:2018ab,Ekblaw:2018aa}. Figure~\ref{bucky} shows the structure comprising  flat tiles that move in space to self-assemble into the ball-like structure. The prototype uses magnetic joins to join the tiles.  Each tile is essentially a sensor node, designed to communicate with other tiles (e.g., via Bluetooth) and a base station. Sensors in each tile include Inertial Measurement Units (IMUs), Hall sensors, Light Detection and Ranging (LIDAR) Time-of-Flight sensors and a microphone. These tile-embedded sensor tiles could be used to create radiation detection systems, life support monitoring systems and adaptive changes to the structure. 

While such self-assembling structures have been demonstrated on a small scale, one could envision large scale structures being constructed in this way, each part not just a sensor node but equipped with autonomous sensing and reasoning capabilities and actuators (effectively, a robot) that could find, position, and connect physically with other parts to form large structures, even buildings or stations in space. 

Hence, we could see different types of robots swarms (or smart parts, e.g., smart tiles as mentioned above) that communicate with each other and  cooperatively self-assemble, themselves being part of a structure, rearranging on-demand to adapt when needed, or swarms of constructor robots that are connected to each other and to base stations and can collaboratively construct structures in space or on planets. The notion of robots that construct other robots that do different types of construction is still to be explored. Also, the ways in which such robots communicate, whether via the environment in a stigmergic manner or directly via short range communications, will need further exploration. Such automatic self-assembly is particularly useful when large structures are hard to move from earth to space and in harsh environments where it is not possible for a large number of human workers to work over long periods. 

Various kinds of purpose-built space stations and living environments that might be constructed for space, for example, space hotels for tourism, energy harvesting devices or stations~\cite{graczyk2017energy,7393105,Snowden:2019aa} to space farms~\cite{NASA:2019aa} can be completely automated and will make use of IoT sensing and data processing for remote monitoring of their functions and control. Indeed, methods on earth learnt from IoT-based automation and control would need to be developed further for space.

\subsection{Data Centres in Space and Data Management Services for In-Space Operations} \label{sec:data_centres}

As human activity and operations extend into space, there is a need for compute and storage resources~\cite{Brounstein:2019aa}. Although devices in space can communicate via satellites to ground stations on earth, there are latency and delay issues, and possible issues with transmission over many nodes and distances. Moreover, data centres in space can be connected to devices on earth. Cooling for such data centres is also then a given, being in space.

One could envision data centres, of varying sizes, perhaps starting with micro-data centres in space. This is already being pursued by several startup companies, as the cost of sending things to orbit continues to decrease and the space required for computing power and memory continues to decrease (at least within the limits of Moore's law and its slowdown)~\cite{Donoghue:2018aa}. 

Building data centres to operate in cold environments is still to be further explored. It is noted that the computers in the International Space Station need to back up data often, due to a higher failure rate while in space  due to radiation~\cite{Backup4all:2017aa}. Microsoft has studied data centres to operate deep in the ocean~\cite{Microsoft:2020aa}. Microsoft’s Project Natick team deployed a data centre 117 feet deep on the seafloor in 2018 for two years. It was noted that  servers in the underwater data centre were eight times more reliable than data centres on land. Similar experiments are yet to be done for large scale data centres in space.

The work in~\cite{DBLP:journals/symmetry/PeriolaAO20} proposed data centres in space (in the Earth’s orbit) which can make use of water mined from asteroids  for cooling  in order to reduce the usage of Earth’s water for data centres. This is argued to be possible due to water bearing asteroids coming near earth, once a year. Such data centres or compute farms in space would need to be automatically maintained or their management automated, if not at least be remotely operated. 

We discussed edge computing for space earlier in Section~\ref{sec:edge_computing}. Indeed, it is not just storage but performing computations in space is needed (e.g., to reduce data storage by storing processed or summary data and perhaps discarding much of raw data, where possible, and to perform on-board processing instead of data transmission to earth for processing). Already, satellites have computers and the idea of satellites with Graphics Processing Units (GPU) and processing power has been considered in order to perform deep learning inference~\cite{Kothari:2020aa}. There could be satellites or space stations that are dedicated to compute servers for other satellites that are more limited in computational capacity or specialized for other functions. There has been recent work exploring machine learning on data and inference in space, on resource-constrained low-power devices (e.g., small satellites) - also called Orbital Edge Computing (OEC)~\cite{Denby:2019aa,lofqvist2020accelerating,10.1145/3373376.3378473} - including using groups of such satellites.

Large cloud service providers are beginning to provide data services for space. For example, the Amazon Web Services (AWS) Ground Station links up satellite dishes to  cloud services and enables data downloads  from satellites.\footnote{\url{https://aws.amazon.com/ground-station/}} The Microsoft Azure Space project\footnote{\url{https://news.microsoft.com/azurespace/}} is working on connecting Azure datacenters on earth to satellites, and providing Azure cloud services for space. Efforts to inter-connect the networks shown in Figure~\ref{fig:Overview} will  be important for future space data services.
  
There are also possibilities for inter-planetary cloud data centres. For example, if and when a Moon base or a base in Mars is set up, data centres will need to be set up on these  planets and such data centres will also need to link up with data centres in satellites (orbiting earth, Moon and Mars or other planets) and data centres on earth. The inter-planetary Internet to enable such links has been discussed at least as far back as 2005~\cite{Jackson:2005aa} with deep space Internet tests as early as 2008~\cite{NASA:2008aa} and more recent discussions on architecture designs for the deep space Internet~\cite{8799061}. The delays in transmitting data between planets (and satellites) are much longer than between points within the same planet (e.g., Earth and Mars transmission can take between three to twenty-two minutes at the speed of light) so that DTN approaches are needed. Practical deployment of accessible deep space testbeds will need to be developed for research in this area.

\subsection{Robots in Space} \label{sec:robots}
Apart from large robot arms for assembly or self-assembling parts, there could be robots helping humans within space stations. Kirobo is Japan's first robot astronaut, which was demonstrated in the International Space Station in 2013. But it is used only in a limited way - e.g., to entertain human astronauts, and had spent eighteen months in the ISS~\cite{Dentsu:2015aa}.

\begin{figure}[hbt]
\begin{center}
\includegraphics[width=90mm,height=45mm]{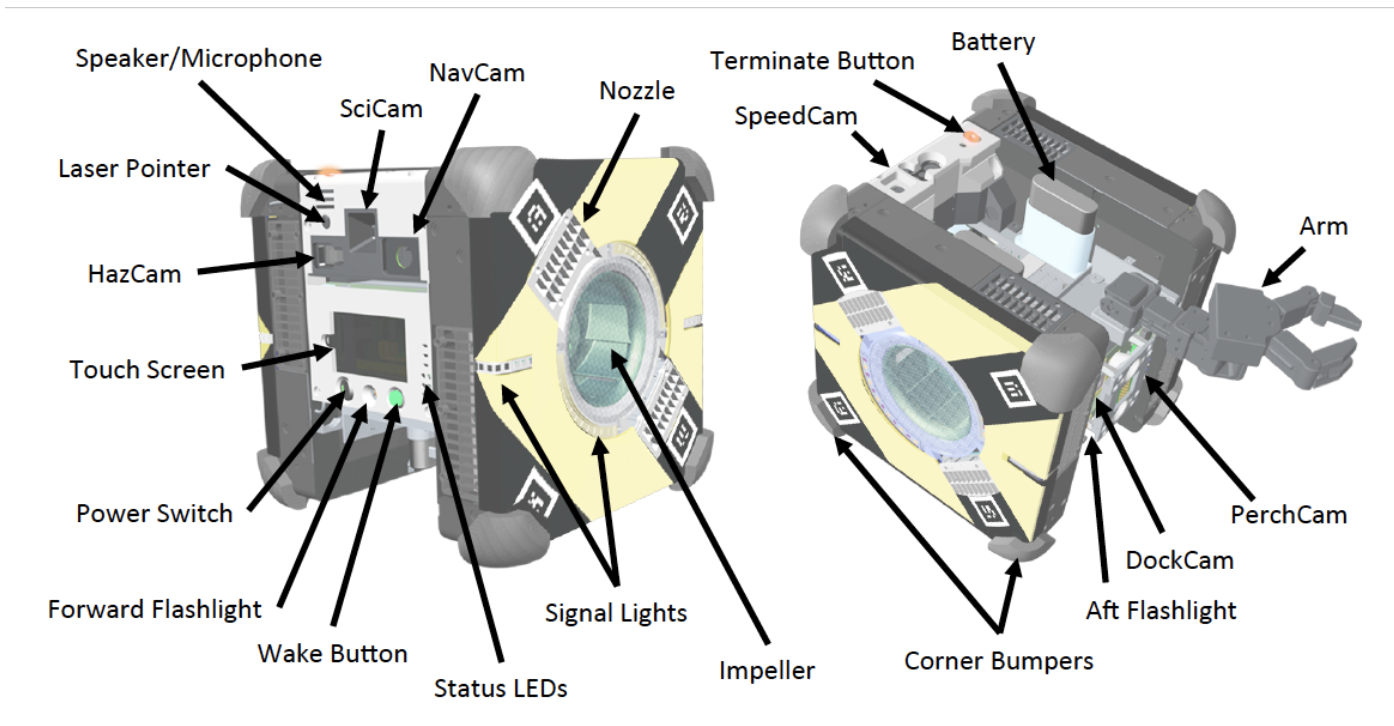}
\caption{Astrobee Architecture (From~\cite{Bualat:2018aa}, Credits: NASA, https://www.nasa.gov/astrobee)}
\label{Bualat:2018aa}
\end{center}
\end{figure}

An interesting deployment of robots to help astronauts are NASA's cube-shaped Astrobee robots, each around 12.5 inches wide.\footnote{\url{https://www.nasa.gov/astrobee}} These robots are free-flying and can help astronauts in their daily routine work, including helping astronauts move cargo throughout a space station and documenting experiments using their built-in cameras. They use small arms to grip and toss themselves from one rail to another. Each Astrobee is equipped with a speaker/microphone, a laser pointer, a touch screen and lights (e.g., status LEDs) for user interaction~\cite{Bualat:2018aa}. Figure~\ref{Bualat:2018aa} illustrate the components of Astrobee.

Swarms of such robots could inhabit space stations (or space buildings), or space vehicles (or spaceships), of the future, and help humans with a wide range of tasks. Such robots could potentially move from one station to another in performing their tasks, from repair to maintenance. On other planets, rover robots are well known, e.g., the Mars Rover,\footnote{\url{https://mars.nasa.gov/mer/}} which can spend years in operation on the planet.  Other types of robots have been explored \cite{Woolastone:2020aa}.

Technical open challenges of space robotics have been identified~\cite{IEEE:2020aa}, including manipulation of objects in zero or micro-gravity conditions, mobility in tough and rough planetary environments different from earth (requiring advanced sensing and perception, and mechanical agility), effective teleoperation (and human-robot interaction) and adjustable autonomy. Indeed, such challenges for using robots for space exploration have been noted in~\cite{Gaoeaan5074}. 

In relation to IoT, and the recent notions of the Internet of Robotic Things (IoRT), there is a focus on connectivity as central in robots coordinating and functioning, e.g., the use of cloud/fog/edge-enabled robots and networked robots. For such robots in planets and space, a challenge is how such robots can be empowered by cloud/fog/edge resources - for example, a connected planetary rover that can use additional storage or computational resources in a ground-based data centre and/or in a satellite orbiting the planet. There is also the challenge of how such robots can be remotely managed and maintained (e.g., software updates) and how a massive swarm of such robots can be coordinated. The necessary infrastructure to maximise the effectiveness of such robots is an issue - e.g., localization/positioning infrastructure, in addition to cloud resources.  Such infrastructure must also be self-maintaining and self-healing given that it would be hard or impossible for human workers to attend to them.

\subsection{Connected Automated Space Vehicles} \label{sec:connected_vehicles}

As companies continue to build reusable, rather than disposable, vehicles to bring people to space, cargo and deliveries, and to spend time in space (e.g., for tourism), there are other types of vehicles that could be  developed, in time, for use in space. For example, automated space-ships of different sizes and capacities, for short or long distances, allowing travel from one space station to another is another idea; there are not many human habitable space stations in orbit at this point, and so, this is currently not so useful. However, future inter-station travel in space might require such vehicles. 

Traffic management in space will be required. Dealing with space debris is only one problem (discussed later), albeit a significant one, but the need to create space highways (akin to air highways proposed for drones in the sky~\cite{BBC:2020aa} might arise to regulate and manage space traffic  (in three dimensions in contrast to terrestrial road traffic management). There would be considerable issues from licensing and unique identification of space vehicles, to ways and means for zoning in space - issues perhaps too far in the future to consider further here. 

Vehicle-to-vehicle connectivity and inter-vehicle sensing will be crucial aspects of technologies in such vehicles for space, to sense one another to avoid collisions but also to identify and communicate with each other, and to cooperate in their movements. 
 
Similar issues can be considered in vehicles (e.g., rovers) on other planets, e.g., on the moon. However, it would require a large number of such vehicles before traffic management of vehicles on a planet needs to be managed.

\subsection{Networked Wearables and Apps in Space} \label{sec:networked_wearables}

As far back as the late 1950s, NASA had pioneered devices to 
measure an astronaut's temperature, respiration, and cardiac activity \cite{NASA:2006aa} Such techniques were later adapted for patient monitoring in hospitals. 

There have been continued work on health monitoring devices for space travel, e.g., as discussed in~\cite{Rafiq2015}, the Bio-Monitor \cite{CSA:2019aa} France’s \R{National Centre for Space Studies (CNES; French: \textit{Centre national d'études spatiales})} app for astronauts to use on a tablet device called EveryWear (which integrates health and medical related functions) and wearable garments with bio-sensors (such as Astroskin evaluated by NASA~\cite{NASA:2015aa}. As far back as 2013, brain computer interfaces for space travel have  been proposed and discussed~\cite{warwick13}. Understanding the physiological and psychological effects, short and long term, on the human space traveller is an intensive area of research, broadly known as Bioastronautics~\cite{EVETTS2009105,stwlzjz20}. Different types of wearable technologies could be used in space, ranging from health monitors, robo-gloves~\cite{NASA:2012aa} spacesuits (with embedded sensors for temperature, pressure, gas, and humidity, biosensors, and computers) to exoskeletons, \R{e.g.,} \cite{NASA:2018ab} and GoPro-type cameras. Recent work has seen garment-integrated wearable biosensors with e-textiles, with sewn-in electrodes  for heart rate monitoring~\cite{arquilla21}, and wearable biosensors to measure the biological clock neuropeptide, orexin/hypocretin, in sweat~\cite{amber21}.

New types of wearables with actuators, apart from sensors, from wearable robots to a range of wearable assistive prosthetics might be considered, e.g., the  wearable soft-robotic additive prosthetic resembling a seahorse tail in~\cite{sumini21} that can support  astronauts during Extra Vehicular Activities, and a wearable soft pneumatic device to stimulate gait muscles during lunar missions~\cite{ticllacuri21}.

Apart from essential health monitoring, envisioning recreational space travel in the future, passengers (e.g., space tourists) would want similar access to their smartphone and smartwatch apps as they would have on earth, of course, with adaptations. There are numerous questions in relation to how to support computational, storage and networking needs of space travellers - e.g., if people spend months in space, then they will need to have their own social network and photo sharing services as well as online workspace, and such services  might synchronise (though with delays) with earth-based sites. Such wearables will also need to connect with edge and fog computing data centers in space and platform services in space as we discussed earlier. 

Augmented Reality (AR) and Virtual Reality (VR) applications might be desired, which require connectivity with adequate bandwidth and low enough latency. Allowing remote user friendly interactive experiences via virtual telepresence environments in space is a NASA space technology grand challenge \cite{NASA:2015aa} -- for participants from earth and space (and even other planets) to meet in the same virtual environment with near real-time interaction, adequate networking infrastructure spanning vast distances would be required -- we also discussed the challenges of inter-planetary networks earlier. As an example, the work in~\cite{zeidler21} explored the use of augmented reality to support workers for greenhouses in Mars or Moon, e.g. to display plant identification information and to support communication and planning.

\subsection{Space Situational Awareness, Dealing with Space Debris and Space Traffic Management} \label{sec:situational_awareness}

Space debris, with tens of thousands of space objects~\cite{Falk:2014aa} are problematic if space is to be further populated. Massive cooperation among different parties are required to be able to adequately track objects in space to ensure the safety of space flights\footnote{See initiatives such as \url{https://www.space-track.org/documentation\#odr} and \url{https://spaceorbits.net}, and the satellite map at \url{https://maps.esri.com/rc/sat2/index.html}} - space situational awareness is required if collisions are to be avoided and the space within the earth's orbit is to be best utilised. Machine learning techniques can be used to improve orbit predictions, building on physical models, where data is not complete. For example, see the Space Situational Awareness in Low Earth Orbit project by IBM.\footnote{\url{https://github.com/IBM/spacetech-ssa}}

While there have been extensive IoT applications in tracking people, animals, and things, as well as events and situations at places, the challenge of tracking people and things in space will be enormous, not only in detecting and tracking space debris but in tracking people and things as they are increasingly being sent into the earth's (low, medium and high) orbits. The Space Fence system~\cite{Eversden:2020aa, DOTE:2019aa} which has recently come online, is used to track space debris and satellites in space. It is an S-band radar system and can detect small objects not previously tracked. In general, new radar systems and localization systems within 3D space will be required, e.g., reference stations in space, radar-based methods or inertial navigation type methods, for vehicles that move in space as well as finding objects and stations in space such as in inter-station travel.

Within space stations, there is also a need to avoid congestion among astronauts and to locate astronauts and equipment -- hence, positioning technologies to locate humans and their status, and to locate equipment and tools, within space stations is required~\cite{marquez21}. 

\subsection{Colonising Planets} \label{sec:colonising}

Recently, NASA has requested Nokia to set up a 4G network on the moon~\cite{ABC:2020aa}. While still in the imagination, such ideas suggest new horizons of infrastructure development on other planets, including infrastructure for sensor networks, navigation (or localisation), and data communications that will be needed wherever humans live. 

There is no GPS on other planets (yet). A planetary rover can navigate by computer vision techniques and image comparisons,\footnote{For example, see the surface perspective to satellite perspective matching approach in 2018: \url{https://frontierdevelopmentlab.org/2018-localization}, and its feature in~\cite{IEEE:2018ab}} or use its own sensors on the  vehicle itself to keep track of the vehicle's movements - e.g., accelerometers, gyroscopes and wheel odometers, i.e., inertial navigation~\cite{Kilic2019}. Other solutions have been proposed in connection with colonising Mars~\cite{zubrin11}.
An approach using antenna based reference stations can also be used~\cite{PLANS:2018aa}.

There is also a number of interesting projects that have been funded by national space agencies to develop technologies needed to support extra-terrestrial habitats. For example, the NASA-funded Resilient Extra-Terrestrial Habitats (RETH) Project by Purdue University \footnote{\url{https://www.purdue.edu/rethi/}} aims to develop the technologies needed to establish extra-terrestrial habitats, focusing on three specific themes:

\begin{itemize}
    \item System Resilience: Developing technologies to establish resilent systems and computing capabilities to make decisions regarding habitat architecture and onboard decisions;
    \item Situational Awareness: Developing robust and automated methods for detecting and diagnosing system faults;
    \item Robotic Maintenance: Developing methods to realize teams of independent autonomous robots to navigate through dynamic environments, and perform collaborative tasks.
\end{itemize}

Up to this point, we have already discussed the use of exploration robots on planets such as the wheeled rover robots. More sophisticated transportation land and air infrastructure will be needed to accompany human-habitats on planets. Futuristic endeavours such as making planets habitable remains an open problem. Elon Musk's idea that humans could live on Mars by 2060s~\cite{Drake:2016aa} will need to be supported by much research on how this can be done systematically. Recent mathematical modelling~\cite{salotti20} estimated that the minimum number of settlers for survival on another planet such as Mars is 110 individuals - the minimum viable population - considering the need for humans to work and produce what is needed for survival (e.g., in order to build and work systems to acquire water, oxygen, and power) - hence, it is not something that can be done by a small group of persons. Such a population of 110 people \hbox{will need to be technologically supported.}

There are questions, given one starts from scratch on a new planet, with today's technology, of what would be the required digital and physical infrastructure to be laid out, and in what order, and how  this will be managed - e.g., basic required infrastructure could be laid out first via self-assembly and automation before the first humans arrive, or should the development of infrastructure proceed hand-in-hand with human arrivals. For example, it could be first  the deployment of living quarters with the basics required for human living, e.g., water, energy, food production, power and oxygen, basic computational and data communications,  with adequate automation (e.g., self-monitoring and self-healing devices) to reduce the burden of system maintenance; and then followed by additional deployment of infrastructure for better connectivity and communications, and thereafter,  further deployment of connected/networked automation systems, robots and devices to perform tasks, controllable and manageable by humans via data connectivity - such data connectivity also helps  inter-device coordination and functions, including navigation and software updates, e.g., mesh networks can be used for data communications and localization, in the case where an infrastructure of base stations has not yet been setup.  

How to build a colony on another planet has been considered in detail in~\cite{Ceriotti:2016aa}. Beyond technological questions, are questions of required human culture and ethics, and social and psychological factors, on a new planet~\cite{SZOCIK2020102514,SZOCIK2020101388,mars19}, e.g., deep altruism is required as well as the right culture, for human survival - these could be mediated by the right technology, e.g., for law enforcement and monitoring (within ethical restrictions).

\subsection{Advances in Satellite Communications} \label{sec:advances_satcom}

Satellite systems have grown exponentially more complex since their inception. In the coming years, they are going to grow much more complex with the recent advances in information technology, telecommunications, and more importantly the world of \R{ubiquitous intelligence powered by AI}. The future of satellite systems is headed towards more complex payloads with the ability to dynamically manage the satellite capacity, using techniques such as beamforming, optical communications, ability to automatically manage multi-orbit constellations wherein the inter-satellite and inter-orbit links are established over geosynchronous and non-geosynchronous satellites. In the context of IoT applications, the most important advances are being made towards setting up 5G-satellite ecosystem~\cite{Sharma:2018aa}, inter-planetary communications wherein satellite swarms will be deployed over other planets~\cite{Farrag:2019aa}. 


\subsection{Summary and Discussion}
 
This section has reviewed and discussed open challenges and futuristic scenarios including smart architectures and construction in space, data centers and computational resources for space (or space cloud computing), the idea of connected robots in space, the notion of automated space vehicles, the idea of wearables for space travel and space apps, space situational awareness and tracking objects in space, infrastructure aspects of colonising planets, and future advancements of satellite communications.  They are inter-related but have the challenges of connectivity, automation, scaling over vast distances, and adaptations to the hostile environment of space.

Table~\ref{tab:future} summarises our discussion of technologies and challenges for each topic in this section - the challenges suggest avenues of future research and development in space initiatives, many of which involve IoT related technologies and networked smart things - no doubt, future work will also see new types of technologies this review has not surveyed.

\section{Conclusions} \label{Conclusions}

The emergence and rapid \R{technological developments in IoT, satellite-based non-terrrestrial/terrestrial communications technologies, edge/fog/cloud computing, large-scale data processing powered by AI/ML capabilities} have enabled many opportunities for space exploration. Many of these technologies have evolved over time, and as they mature, it has become increasingly evident that IoT can be seamlessly integrated with space-based technologies for further space exploration. Ideas such as space tourism and Mars travel that were once deemed futuristic are no longer improbable. 

In this paper, we have reviewed \R{the state-of-the-art and recent developments in IoT and the space industry, spanning from innovations in the multi-layer non-terrestrial and terrestrial networks, to communications and computing capabilities of various components an integrated space, air, ground network architecture. We have identified potential enhancements to current technologies and outlined avenues for future work. We further discuss future opportunities and technological challenges for satellite communications and space exploration enabled in part by \hbox{IoT technologies, and vice-versa.}} 

As communications and computing technologies become more pervasive, we need to keep building on a foundation of knowledge of automation and ubiquitous connectivity. While progress in IoT, AI and communications technologies have proceeded in somewhat haphazard ways in the past decades, there is also an opportunity to build infrastructure with a clean slate -- but how one does that and address its surrounding challenges remains to be explored.

\ifCLASSOPTIONcaptionsoff
  \newpage
\fi
   
\onecolumn
{\small
    \begin{longtable}{|p{0.12\textwidth}|p{0.73\textwidth}|p{0.08\textwidth}|}
    \caption{\R{Summary of the topics, their areas of concerns and research directions (with representative references) as described in Section~\ref{current_dev_challenges}.} \label{tab:current} }
    \centering \tabularnewline
    \hline
    \textbf{Topics} & \textbf{\color{black}Technologies covered, challenges, and research directions} & \textbf{\color{black}References}\\    \hline
    Satellite communications aided IoT applications (Section~\ref{sec:applications}) & 
    \R{Areas of concern:
    \begin{itemize}
        \item Wide area coverage for IoT services for remote connectivity;
        \item Satellite-aided backhaul connectivity for delay tolerant applications;
        \item Achieving low-latency performance via satellite communications for delay sensitive applications.
    \end{itemize}}
    \R{Research directions:
    \begin{itemize}
        \item Innovations in satellite communications technologies to better integrate with IoT connectivity technologies to cover and reach remote, infrastructure-less environments;
        \item Improving backhaul connectivity to support delay tolerant applications, such as in the event of disasters;
        \item Extensive research needs to address the need for low-latency ultra-reliable performance enabled by satellite communications to enable delay sensitive applications, such as autonomous vehicles, industrial automation, etc.
    \end{itemize}} & \cite{Ding:2020aa,Bean:2017aa,Gharanjik:2018aa,Lee:2010aa,De-Sanctis:2015aa,Berioli:2011a,Fraire:2019aa}\\    \hline
    IoT satellite-terrestrial integrated networks (Section~\ref{sec:integratedNetwork}) & 
    \R{Areas of concern:
    \begin{itemize}
        \item Satellite-to-satellite, satellite-to-earth, and satellite-to-earth intermediary communications;
        \item IoT communication protocols such as CoAP, MQTT are designed for general use cases in terrestrial networks;
        \item Non-GEO satellites are sufficient for most IoT services, with some works looking into using ICN to assist with traffic distribution and visibility.
    \end{itemize}}
    \R{Research directions:
    \begin{itemize}
        \item Identification of application requirements for the types of satellite communication support needed, balancing the requirements for wider coverage, throughput, latency, reliability and resiliency;
        \item Optimisation of IoT protocols to support communication between terrestrial and non-terrestrial networks;
        \item New or modified integration layer between non-GEO satellites and ICN (or similar) infrastructure.
    \end{itemize}} & ~\cite{Giotti:18,Soua:2018aa,Jin:19,Siris:16}\\    \hline
    Satellite-based 5G (and beyond) networks for IoT services (Section~\ref{sec:satellite5g}) & 
    \R{Areas of concern:
    \begin{itemize}
        \item Integration of satellite and 5G (and beyond, 6G) networking technologies and architectures;
        \item Optimisation of connectivity technologies and communication protocols;
        \item Dynamic traffic management and resource allocation cater for heterogeneous IoT systems with widely-varying traffic characteristics and requirements.
    \end{itemize} }
    \R{Research directions:
    \begin{itemize}
        \item Research into more seamless integration between satellite communications, IoT and 5G/6G mobile technologies;
        \item Improving remote connectivity with mobile cell sites with LEO satellites, CubeSats for 3D network coverage;
        \item Potential for using a cooperative infrastructure between satellites and base stations in ground networks;
        \item Leveraging advances in SDN, NFV, AI/ML and Blockchain technologies to build an integrated satellite/5G/6G hybrid network.
    \end{itemize}}
    & \cite{Saad:2019aa, Cao:2018aa, Gineste:2017aa, Bontu:2014aa, Fang:2020aa, Evans:14, Liolis:2019aa}\\    \hline
    Architectures and protocols (Section~\ref{sec:architectures})& 
    \R{Areas of concern:
    \begin{itemize}
        \item H-STIN aims to integrate various standalone architectures and wireless communications protocols to achieve an intelligent framework for IoT and space communications;
        \item Miniaturised satellites such as CubeSats aim to provide global connectivity at lower costs, with applications such as aerial reconnaissance, asset and environment monitoring, disaster prevention, etc;
        \item SDN-based CubeSats network for remote sensing, cellular backhaulng and mission critical communications;
        \item UAV equipped with communications and IoT devices to provide services such as sensing, data collection, target identification, etc;
        \item Data transmission largely relies on TCP and the transport layer, with various data formats also being proposed at the application layer to maximise efficiency over expensive satellite links.
    \end{itemize}}
    \R{Research directions:
    \begin{itemize}
        \item Further enhancements to H-STIN with a comprehensive SAG-IoT network that provides an architectural framework for seamlessly integrating space, air, and ground network segments, each with different various traffic/network characteristics, and requirements;
        \item Active research in the flying formation of UAVs, building on the existing centralised or decentralised control to cooperatively support IoT services more efficiently;
        \item Ongoing research on developing and optimising transport protocols for satellite communications (network paths with high capacities and long delays), and more efficient data formats in terms of energy efficiency, transmission efficiency, and the impact of operating in unlicensed frequency bands.
    \end{itemize}}
    & \cite{Chien:2019aa,Saeed:2020aa,Kak:2018aa,Akyildiz:2019aa,Liu:2018aa,Kato:2019aa,Hong:2020aa,Wang:2009aa,Caini2007aa,Akyildiz2001aa,Lysogor:2018aa,rfc2760,rfc8975,kuhn2020quic,kuhn2020quic,wang2018performance,jones-tsvwg-transport-for-satellite-00} \\    \hline
    Edge computing with satellites (Section~\ref{sec:edge_computing}) &
    \R{Areas of concerns:
    \begin{itemize}
        \item Satellite-enabled Internet at low earth orbit level with STN;
        \item MEC to improve QoS of STN mobile users, andcheduling of data exchanges between satellites and earth stations;
        \item STECN architecture uses LEO satellite networks with hierarchical and heterogeneous edge computing layers/clusters to service user requests.
    \end{itemize}}
    \R{Research directions:
    \begin{itemize}
        \item Investigation of techniques to provide ubiquitous, high data rate and reliable satellite-enabled Internet connectivity, in particular for users in remote areas where terrestrial communication infrastructure is lacking;
        \item Optimise computational offloading techniques to support MEC for STN mobile users;
        \item Develop an architecture for resource pooling of multiple MEC servers within the coverage of LEO satellites.
    \end{itemize}} & \cite{Hu:2001aa,Zhang:2019aa,Spangelo:2015aa,Marinelli:2011aa,Jia:2017aa,Xie:2020aa,Wang:2019aa}\\    \hline
    \end{longtable}
}
\twocolumn

\onecolumn
{\small
    \begin{longtable}{|p{0.12\textwidth}|p{0.73\textwidth}|p{0.08\textwidth}|}
    \caption{\R{Summary of the topics, their areas of concerns and research directions (with representative references) as described in Section~\ref{emerging_dev_challenges}.} \label{tab:future}}
    \centering \tabularnewline
    \hline
    \textbf{Topics} & \textbf{\color{black} Technologies covered, challenges, and research directions}  & \textbf{\color{black} References}\\    \hline
    Smart architecture and construction in space (Section~\ref{sec:smart_architecture}) & 
    Areas of concern:
    \begin{itemize}
        \item Automating assembly tasks via large robotic arms;
        \item Swarm robot assembly; self-assembling structures;
        \item Specialised structures, including required automation and sensing systems, e.g., for agricultural space farms, space hotels.  
     
    \end{itemize} 
       {\color{black} Research directions: 
       \begin{itemize}
           \item Investigation of what structures can be feasibly and efficiently assembled in space using large robotic mechanisms (e.g., robotic arms) versus using swarm robot assembly mechanisms;
           \item Design of specialised structures in space parallelled with specialised buildings and places on earth.
       \end{itemize}
       }
       & ~\cite{Nagpal:2002aa, Chandler:2019aa,Ekblaw:2018ab,Ekblaw:2018aa} \\    \hline
    Data centres in space and data management services for in-space operations (Section~\ref{sec:data_centres})&
    Areas of concern:
    \begin{itemize}
        \item Reliability of in-space data centers and in-space compute servers;
        \item Automated management of in-space data centers and in-space compute servers;
        \item Connecting in-space data centers to terrestrial stations;
        \item Smart energy management for in-space computations and storage;
        \item Issues of communication latency and reliability for inter-planetary cloud data centers.
    \end{itemize} 
          {\color{black} Research directions: 
       \begin{itemize}
       \item Methods to automate management of in-space data centers that will run over long periods of time;
       \item Robustly network in-space data-centers to terrestrial stations;
       \item Ways to sustainably power in-space data centers. 
       \end{itemize}
    
    }
    & \cite{Donoghue:2018aa,DBLP:journals/symmetry/PeriolaAO20,Denby:2019aa,lofqvist2020accelerating,10.1145/3373376.3378473,NASA:2008aa,8799061} \\    \hline
    Robots in space (Section~\ref{sec:robots}) & 
    Areas of concern:
    \begin{itemize}
        \item Robot function and mobility in micro-gravity conditions;
        \item Sensing and perception for robots in space;
        \item Effective teleoperation (e.g., from earth to space and other planets);
        \item Infrastructure for functioning of robots in space and other planets, e.g., positioning/localization and edge/cloud resources to augment robots.
    \end{itemize} 
       {\color{black} Research directions: 
    \begin{itemize}
       \item Identifying applications of robots in space and developing robots that function in space environments; 
       \item Developing the right infrastructure for robots to function in space (within stations and outside), e.g., the need for robots to be tracked and to support them in performing compute-intensive tasks.
    \end{itemize}
    
    }
       & \cite{Bualat:2018aa,Woolastone:2020aa,IEEE:2020aa,Gaoeaan5074} \\    \hline
    Connected automated space vehicles (Section~\ref{sec:connected_vehicles}) & 
    Areas of concern:
     \begin{itemize}
        \item  Reusable space vehicles with smart technology for navigation and maintenance; 
        \item Inter-``space vehicle'' communications and cooperative behaviours;
        \item Traffic management and routing in space.
     \end{itemize}
      {\color{black} Research directions: 
       \begin{itemize}
       \item Identifying a range of transport modes in space, e.g., single person transportation across stations or multiple person transportation, and what the ranges are of these transportation modes;
       \item Traffic management of space vehicles - the vehicles move in 3D free-form space, and so, collision avoidance sensing as well as suitable routing is required, especially in busy regions of space.
            \end{itemize}
    
    }
    &  \cite{BBC:2020aa} \\    \hline
    Networked wearables and apps in space (Section~\ref{sec:networked_wearables}) & 
    Areas of concern:
         \begin{itemize}
        \item Smart sensors and devices for health monitoring for people in space;
        \item Wearable devices to aid space travellers and astronaut tasks; 
        \item Access for people travelling in space to digital/Internet services typically available on earth (e.g., embedded Web servers in space);
        \item Augmented reality, cross-reality, and virtual reality applications for space travellers and operations.
     \end{itemize}
      {\color{black} Research directions: 
       \begin{itemize}
       \item Identifying the wearable devices used by different categories of space travellers; e.g., tourists and astronauts on specific missions; a question is will general purpose wearable devices (akin to smartphones) be developed for space travellers?
       \item How will digital services be supported on such wearable devices, akin to the Internet we have on earth; a question is how will wearable devices in space access content from other earth and space servers? (and what sort of connectivity and bandwidth will be available for space travellers?) One can imagine a space tourist travelling from earth to the moon wanting to check emails - will this be possible?
       \item We can imagine new applications of AR/VR in space - a question is how such functionalities will be supported?
       \end{itemize}
    
    }
      & \cite{Rafiq2015,NASA:2018ab,warwick13,EVETTS2009105,stwlzjz20,NASA:2015aa,sumini21,ticllacuri21,zeidler21,arquilla21,amber21} \\    \hline
    Space situational awareness, and dealing with space debris (Section~\ref{sec:situational_awareness}) & 
    Areas of concerns:
        \begin{itemize}
        \item Technologies for real-time situational awareness in space to avoid space debris;
        \item Detection and localization of space debris and clean-up;
        \item Situational awareness and localization within space stations;
     \end{itemize}
      {\color{black} Research directions: 
       \begin{itemize}
      \item Space debris and cleaning up remain challenging problems to solve; while there is space debris, the need to avoid such debris  when moving through space is required - either to detect and navigate around them, or to eliminate them on the way;
      \item Develop techniques for space stations to localise and track their positions relative to other space stations, especially, if movements occur often;
      \end{itemize}
    
    }
     & \cite{Falk:2014aa,Eversden:2020aa, DOTE:2019aa} \\    \hline
    Colonising planets (Section~\ref{sec:colonising}) & 
    Areas of concern:
         \begin{itemize}
        \item Telecommunications and global positioning infrastructure on other planets;
        \item Extra-terrestrial smart habitats (advanced versions of today's smart homes);
        \item Smart mobility and mobility services) on other planets;
        \item Energy, water, food, and oxygen supply and generation (with required automation and sensing systems) on other planets.
     \end{itemize}
      {\color{black} Research directions: 
       \begin{itemize}
       \item Colonising planets remains a far-into-the-future prospect and the complexity of issues cannot be outlined just in this paper and would go beyond IoT research and development; however, we note here the issues of how to provide physical habitats and physical infrastructure on a planet as well as the digital infrastructure (e.g.,  GPS and telecommunications infrastructure on other planets to support IoT applications on other planets, which may have different atmospheric conditions and properties (e.g., gravity)).
       \end{itemize}
    
    }
     & \cite{ABC:2020aa,zubrin11,Kilic2019,PLANS:2018aa,IEEE:2018ab,Drake:2016aa,salotti20,SZOCIK2020102514,SZOCIK2020101388,mars19,Ceriotti:2016aa} \\    \hline
    Advances in satellite communications (Section~\ref{sec:advances_satcom}) & 
    Areas of concern:
         \begin{itemize}
        \item 5G/6G-satellite ecosystem; 
        \item Inter-planetary (and deep space) communications;
        \item Satellite swarms over other planets.
     \end{itemize}
      {\color{black} Research directions: 
       \begin{itemize}
       \item A Space Internet that crosses a range of networks shown in Figure~1 remains a challenge - e.g., can someone under the sea on earth communicate with someone walking on the moon? (this itself would be a challenge in connectivity but already feasible by connecting networks; further developments towards robust inter-network connectivity will be needed);
       \item The size and scale of satellite swarms, and their applications, continue to be areas of active development, not only by researchers but also in industry.
      \end{itemize}
    }
     & ~\cite{Sharma:2018aa,Farrag:2019aa} \\    \hline
    \end{longtable}
}
\twocolumn

\bibliographystyle{IEEEtran}
\bibliography{iot_space_paper.bib}

\end{document}